\documentstyle[12pt]{article}
\begin{document}

\begin{center}
Submitted to {\em FEBS Letters}, June 1994\\
\vspace{1pc}
{\Large {\bf Protein-Machine model of a single enzymatic reaction}}\\
\vspace{1pc}
{\bf Micha{\l} Kurzy\'{n}ski}\\
\vspace{1pc}
Institute of Physics, Adam Mickiewicz University,\\
Matejki 48, PL-60-769 Pozna\'{n}, Poland\\
fax: (61)-658-962,
e-mail: kurzphys@plpuam11.amu.edu.pl
\end{center}

\vspace{1pc} \noindent{\bf Abstract.} Formulation of a truly advanced
statistical theory of biochemical processes needs simple but realistic models
of phenomena underlying microscopic dynamics of proteins. Many experiments
performed in the 1980s have demonstrated that within the protein native
state, apart from usual vibrational dynamics, a rich interconformational
(activated) dynamics exists. The slowness of this dynamics makes any
conventional theory of chemical reactions inapplicable for description of
enzymatic reactions. It is presumably a rule that it is the process of
conformational relaxation, and not the details of chemical mechanism, that
affects their rate. In a simple model of Protein-Machine type, applied in
constructing a novel theory of enzymatic reaction, conformational dynamics is
treated as a realative quasi-continuous motion of solid-like structural
elements of protein. Simple and tractable formulas for the chemical
relaxation time and the enzyme turnover number in the steady state conditions
are found. The important result obtained is that the kinetic mechanisms close
to and far from the equilibrium can differ.

\vspace{2pc}

\newpage
\noindent
{\bf 1. INTRODUCTION}
\vspace{1pc}

Any statistical theory of physical processes has to refer to simple models of
phenomena underlying microscopic dynamics. The transition state theory,
commonly used in interpretation of biochemical reactions [1], assumes that
biomolecules, in particular enzymatic proteins, perform only fast vibrations
about a single well-defined equilibrium conformation. This picture, adapted
directly from chemistry of low-molecular weight compounds, has proved,
however, untrue. Many experiments performed in the 1980s using various
techniques have demonstrated that within the protein native state, apart from
usual vibrational dynamics, a rich interconformational (activated) dynamics
exists in the whole range of time scales from $10^{-11}$s to $10^5$s or
longer [2, 3]. The slowness of this dynamics makes any conventional theory of
chemical reactions inaplicable for description of biomolecular reactions [4].
A consequence is a challenge to physicists-theoreticians to construct a truly
advanced statistical theory of biochemical processes based on still simple,
but {\em realistic} models of microscopic dynamics of proteins.

Conformational transitions within the protein native state take place not in
the whole bulk of protein but are limited to liquid-like regions surrounding
solid-like fragments of secondary structure (Figure 1). It is not easy to
infer the actual nature of conformational dynamics from experimental data so
that the problem of formulating appropriate models is to some extent still
left open to speculation. In two classes of models provided hitherto by
literature the speculative element seems to be kept within reasonable
limits.  We refer to them symbolically as {\em Protein-Glass} and {\em
Protein-Machine} [3].

The relaxation time spectrum of conformational dynamics of protein looks
practically like quasicontinuous. The simplest way to tackle such problems is
to assume that the dynamics of a system  is in every time scale alike, i.e.,
the spectrum of relaxation times has a self-similarity symmetry. This
assumption is the core of any Protein-Glass model. Time scaling can originate
either from a hierarchy of interconformational barrier heights or from a
hierchy of bottlenecks met by structural deffects diffusing across the
liquid-like region of protein [2, 3].

The Protein-Glass models behave unrealistically both in the limit of very
short and very long times, and in practice should be resctricted only to a
few levels of the hierarchy [2]. An alternative free of such disadvantages is
the Protein-Machine class of models in which the variety of conformations
composing the native state is assumed to be labelled with a few
``mechanical'' variables, e.g. angles describing mutual orientation of rigid
fragments of secondary structure or larger structural elements (Figure 1).
The Protein-Machine models not only display correct asymptotic behaviour but
can also bring simple and tractable formulas for phenomenological parameters.
It is the aim of this letter to present a few such formulas for a single
enzymatic reaction. A formal derivation of equations will be given in a more
detailed paper now in preparation.

\vspace{2pc}
\noindent
{\bf 2. FORMULATION AND EXPERIMENTAL JUSTIFICATION OF THE MODEL}
\vspace{1pc}

The concept (and the name) ``Protein-Machine'' was proposed for the first
time by Chernavsky, Khurgin and Shnol in 1967 [5] but similar picture of
protein dynamics has been considered independently also by several other
authors [6-10]. In 1982, the model was formalized by Shaitan and Rubin [11]
in terms of diffusion in an effective potential along the mechanical
coordinate. This kind of dynamics with the simplest, parabolic potential has
been applied for interpretation of particular protein reactions by Agmon and
Hopfield [12] and Cartling [13].  A detailed mathematical analysis of the
model applied to a general reversible reaction was performed by the author of
this letter [14].

For the sake of convenience we will assume that the considered mechanical
variable, $x$ (compare Figure 1 for more concreteness), is dimensionless
(proportional to the square root of Bolzmann's constant times the
temperature, $\sqrt{kT}$) and normalized in such a way that its equilibrium
dispersion is 1/2.  Then, the process of diffusion in parabolic potential
with the minimum at $x_0$ is described by the partial differential equation
[14]
\begin{equation} \frac{\partial}{\partial t}p(x,t) =
\frac{\partial}{\partial x}\gamma \left[ \frac{1}{2} \frac{\partial}{\partial
x} + (x-x_0) \right] p(x,t)
\end{equation}
for probability density $p$.
Because of original applications, Equation (1) is named after Smoluchowski or
Ornstein and Uhlenbeck. The second parameter, $\gamma$, has the meaning of
the inverse relaxation time of the mean value of the conformational variable,
\begin{equation} X(t) \equiv \int_{-\infty}^{\infty} dx\, x\, p(x,t) \,.
\end{equation}
Indeed, substitution of Eq.\ (1) into Eq.\ (2) results in the
ordinary differential equation:
\begin{equation} \dot{X} = -\gamma\,(X - x_0)
\end{equation}
(the dot means derivation with respect to time).

A motion of fragments of secondary structure relative to each other is often
being observed when two or more different crystalline structures of the same
protein are compared [15].  Also a relative motion of the whole domains has
been the subject of numerous structural investigations [16, 17]. As is quite
natural, in comparative structural studies only a few, usually two,
conformations are apparent. In the same terms of a few, at the most, {\em
discrete} conformational states, the conformational dynamics is described by
conventional biochemistry [1]. The recent paper by Haran, Haas, Szpikowska
and Mas [18] is therefore important showing clearly, in the study of
fluorescence energy transfer between donor and acceptor centres located on
different domains, a {\em quasi-continuous} distribution of interdomain
distances.

A quasi-continuum of short-lived conformational states within the native
state of protein has been observed directly in numerical simulations [19,
20]. A careful analysis [20, 21] indicated that numerically studied dynamics
of interconformational transitions actually has a character of relative
motion of the secondary structure elements.

It should be noted that the mathematical equation identical to (1) also
describes overdamped collective {\em vibrations} of domains, moreover,
numerical analysis indicated that they also take the form of mutual motions
of relatively rigid fragments of secondary structure [22]. In fact, the
cross-over between sequences of conformational transitions along the
mechanical coordinates and overdamped low-frequency collective vibrations is
more or less a matter of convention as one can introduce {\em effective}
normal modes corresponding to the envelope of the ragged (with many local
minima) potential along chosen directions [23].  Protein-Machine models of
dynamics are universal also in another sense displaying, possibly, a
hierarchy typical for Protein-Glass models. Thus, the mechanical elements can
be identified not only with the fragments of secondary structure, but also
with the larger domains and, on the other hand side, smaller side chains
[24].

A controversion exists as far as the time scale of mechanical motions is
concerned. Direct projection of molecular dynamics trajectories onto the
subspace spanned by the principal axes of the low-frequency normal modes
gives relaxation times of these modes only twice as long as their period,
i.e. 10$^{-11}$ s at the most [25]. This result is in agreement with the
estimations made almost twenty years ago on the basis of a simple
hydrodynamic model [26]. However, in the longest up to date 900 ps ($\sim
10^{-9}$ s) simulation by Amadei, Linssen and Berendsen [27] no equilibration
has been observed of the trajectories projected onto the subspace spanned by
the principal axes of the ``essential'' modes of motion. Also the already
mentioned study of the fluorescence energy transfer [18] indicates that the
motion of domains relative to each other is ``slow on the nanosecond time
scale'' (in fact, the value of the intramolecular diffusion constant and that
of the equilibrium dispersion of interdomain distance from Ref. 18 give the
value of the relaxation time greater than 1 $\mu$s).

{}From what was said above it follows that the mechanical variables, slow by
the definition, cannot be directly identified with the low-frequency normal
modes, but rather with the essential modes introduced in Ref. 27 (the
principal axes of the {\em normal} modes are defined as those in which the
force matrix is diagonal, whereas the principal axes of the {\em essential}
modes, as those in which the covariance matrix of the atomic displacements is
diagonal). Two quoted examples seem to point to a rule that the value of
relaxation time estimated in a given experiment (as such we treat also
numerical simulations) coincides with the value of the longest time
accessible for observation with the help of a given technique. Anyway, our
somewhat hypothetical assumption about the conformational dynamics of
proteins observed in the time scale of enzymatic reactions ($\sim 10^{-3}$s)
[3] having the character of a quasi-continuous mechanical motion does not
contradict the experimental settlements.

\vspace{2pc}
\noindent
{\bf 3. GENERALIZED HALDANE'S KINETICS}
\vspace{1pc}

In conventional biochemistry [1] no distinction is made, in general, between
the kinetic and the chemical mechanism of reaction, and the reaction
involving a single covalent step \[ {\rm R} \longleftrightarrow {\rm P} \] is
usually modelled by the three-step kinetics of Haldane with the only two
association-dissociation steps of the substrate or the product to enzyme
added (Figure 2(a)). However, conformational (non-covalent) transitions are
as slow, if not slower, as the very covalent or binding steps, thus have to
be treated on an equal footing with them. The actual kinetic scheme of a
single enzymatic reaction apears, consequently, infinitely more complex than
the Haldane's scheme. It is shown in Figure 2(b) but it can hardly be
considered a proper phenomenological description of the reaction [4].

In the Protein-Machine model considered, dynamics of conformational
transitions within each of the three distinguished chemical enzyme species E,
ER and EP, is approximated by diffusion along the mechanical coordinate $x$.
Chemical transitions are perpendicular to this coordinate, thus the dynamics
of the enzyme as well as the reaction is described by a set of three coupled
balance equations for probability densities $p_i(x, t)$ ($i$ = 0, 1, 2 for E,
ER and EP, respectively):
\begin{equation} \frac{\partial}{\partial t}p_i =
- \frac{\partial}{\partial x}j_i + w_i ,
\end{equation}
with diffusion fluxes
\begin{equation}
j_i = - \gamma \left[ \frac{1}{2} \frac{\partial}{\partial x} +
(x-x_i) \right] p_i ,
\end{equation}
and local reaction rates
\begin{equation} w_0 = - w' + w'',~~~~w_1 = -w + w' ,~~~~w_2 = w - w''
\end{equation}
(see Figure 2(c)).

We assume that the chemical transitions are localized in narrow regions of
the values of mechanical variable:
\begin{equation} w(x) \propto \kappa
\delta (x) ,~~~~ w'(x) \propto \kappa ' \delta (x-x') ,~~~~ w''(x) \propto
\kappa ''\delta (x-x'') ,
\end{equation}
where $\delta$ stands for Dirac's
delta (the reaction is {\em gated} by conformational dynamics [14]). In more
concrete terms, for one value of the angle $x$ (Figure 1) free space large
enough for the substrate motions inside the enzyme can, for instance, appear.
For another, sharply defined, value of the angle $x$ all the catalytic groups
of the active centre can simultaneously be properly oriented. Yet another
value of $x$ can favour the substrate desorption.

We assume also that local chemical transitions are much faster than
conformational diffusion:
\begin{equation} \kappa , \kappa ', \kappa '' \gg
\gamma .
\end{equation}
This assumption is quite natural if we recall that
conformational relaxation within the native state of enzyme is as fast as the
very reaction so that it is this relaxation that determines the resultant
reaction rate and not the details of the chemical mechanism. The important
consequence of this assumption is in general the difference in the kinetic
mechanism of the reaction close to and far from the equilibrium.

\vspace{2pc} \noindent {\bf 4. REACTION CLOSE TO THE EQUILIBRIUM}
\vspace{1pc}

Close to the total chemical equilibrium three concentrations (molar
fractions) of the free enzyme and its two complexes
\begin{equation}
C_i(t)
\equiv \int_{-\infty}^{\infty} dx\, p_i(x,t)
\end{equation}
($C_0$ = [E], $C_1$ = [ER], $C_2$ = [EP]) evolve towards their equilibrium
values, $C_i^{\rm eq}$, with a single relaxation time, $\tau_{\rm eq}$, given
by the formula:
\begin{equation}
\tau_{\rm eq}^{-1} = \frac{\gamma}{2\sqrt{\pi}}
\sum_{i=0}^{2} (1-C_i^{\rm eq})\sqrt{\Delta G_i^{\ddagger}/kT}\, \exp
(-\Delta G_i^{\ddagger}/kT) .
\end{equation}
Quantities $\Delta
G_i^{\ddagger}$ are values of the free energy that has to be surmounted
within particular species in order to reach the nearest gate from the
equilibrium conformation $x_i$ (Figure 3). Formula (10) is to be derived from
Eqs. (4-7) with the condition (8) and for $\Delta G_i^{\ddagger} \gg kT$
using the variational method [14].

Equation (10) has the meaning of the average reciprocal time of diffusion
uphill the conformational potential from its minimum.  Quite generally, the
time of difusion from $x'$ to $x''$ in the potential $x^2$ (in $kT$ units) is
given by the formula
\begin{equation} \tau (x'\!\!\to\! x'') = 2\gamma^{-1}
\int_{x'}^{x''} dy~e^{\, y^2} \int_{-\infty}^y dx~e^{-x^2} .
\end{equation}
In Figure 4, diffusion times uphill and downhill the potential are plotted vs
the distance $x$ in the logarithmic scale. It is seen that diffusion uphill
can be several orders of magnitude slower than diffusion downhill, taking
place in the time of the order of $\gamma^{-1}$.

On examining Figure 3 we find that after transition in the point $x'$ from
the state E + R to ER, the enzyme can either equilibrate within ER or pass
directly to the next state EP, or even E + P, with the process of
equilibration within the intermediates omitted. As a consequence, direct
component reactions between {\em each}, in general, pair of kinetic states E,
ER and EP, are possible [4]. The rate constant, $k_{ij}$, of the reaction
between species $i$ and $j$ is related to the single relaxation time (10) by
the simple formula [28]:
\begin{equation} k_{ij} = \tau_{\rm eq}^{-1} C_j^{\rm eq}
\end{equation}
(note independence of $k_{ij}$ of the initial species $i$).

\vspace{2pc} \noindent {\bf 5. STEADY-STATE KINETICS}

\vspace{1pc} No partial equilibration within any kinetic species is necessary
if the enzymatic reaction proceeds far from the chemical equilibrium. In
steady state conditions with the concentration of reactant kept constant, [R]
= const, and with the constant removal of product, [P] = 0, the rate of the
reaction is described, as in the conventional approach, by the
Michaelis-Menten law:
\begin{equation} [{\dot{\rm P}}] = \frac{k_c [{\rm R}]}{K_m + [{\rm R}]} .
\end{equation}
The reciprocal turnover number:
\begin{eqnarray} k_c^{-1} &\!=\!& \tau_1(x'\!\!\to\! 0) + \tau_2(0\!\to\!
x'') + \tau_0(x''\!\!\to\! x') \nonumber\\ && + ~(C_1^{\rm eq}/C_2^{\rm eq})
\left[ \tau_2(0\!\to\! x'') + \tau_2(x''\!\!\to\! 0) \right] ,
\end{eqnarray}
and the apparent dissociation constant:
\begin{eqnarray} K_m &\!=\!& k_c
[R]^{\rm eq} \left\{ (C_0^{\rm eq}/C_1^{\rm eq}) \left[ \tau_1(x'\!\!\to\! 0)
+ \tau_1(0\!\to\! x') \right] \right. \nonumber\\ && \left. + ~(C_0^{\rm
eq}/C_2^{\rm eq}) \left[ \tau_2(0\!\to\! x'') + \tau_2(x''\!\!\to\! 0)
\right] \right\}
\end{eqnarray}
($\tau_i$ denoting diffusion time in the
potential of $i$-th species with the minimum at $x_i$) have, however, quite
unconventional interpretation.

The theory presented gives a rule for the optimum action of
enzyme: the turnover number $k_c$ is maximum, if conformational
diffusion within free enzyme E is downhill, and the transition
points $x = 0$, $x'$ and $x''$ lie not very far from each other.
In that case $k_c~\simeq~\gamma~\gg~\tau_{\rm eq}^{-1}$. Of course
the backward reaction along the same path in steady state
conditions should be much slower, but the latter reaction, in
Protein-Machine model, can proceed along another path it may find
more convenient.

\vspace{2pc}
\noindent
{\bf 6. CONCLUDING REMARKS}

\vspace{1pc}
Perhaps the most important result of the presented theory is
that the turnover number $k_c$, Equation (14), of the enzyme in
the steady-state conditions is formally independent of the
reaction rate constants $k_{ij}$, Equations (12) and
(10). In general, the kinetic mechanisms of the reaction close
to and far from the equilibrium can differ. This was suggested
already more that twenty years ago by Blumenfeld [7] and the
content of present paper is, in fact, nothing else than a more
precise formulation of his ideas.

An essential feature of the new approach is that it leaves the
classical phenomenology of the enzymatic reaction essentially
unaltered and changes only interpretation of phenomenological
parameters.  As a consequence, it will be, probably, not simple
to carry out {\em experimentum crucis} directly proving the
conventional theory wrong. Two general predictions made by the
novel theory: (a) the independence of the enzyme turnover number
of a particular chemical mechanism, and (b) a distinct
difference in values of the turnover number for the reaction
proceeding in the forward and backward direction along the same
pathway, do provide the explanation of two undoubtedly specific
properties of enzymatic reactions: (a) a relatively small
dispersion of values of the turnover number about the almost
universal value 10$^3$ s$^{-1}$, and (b) a distict
irreversibility of most reactions in the steady-state
conditions. The reversible reactions observed in these
conditions can be explained by the Protein-Machine theory in
terms of relaxation pathways along two different mechanical
coordinates for the forward and the backward directions.

Only close to the equilibrium can enzymatic reactions be
thermodynamically described in terms of {\em concentrations} of
kinetically distinguishable chemical species. In steady state
conditions far from the equlibrium the proper thermodynamic
variables are rather quantities characterizing conformational
nonequilibrium of the enzyme, that is the {\em mean values of
mechanical variables} in the case of the model considered. The
conformational nonequilibrium should play an extremely important
role in processes of coupling of several reactions taking place
at the same multienzyme complex, the preasumably universal
statistically independent unit of biochemical processes. In the
conventional mechanism of {\em chemical} coupling the complex is
needed only for keeping the appropriately high concentration of
intermediates. The conformational nonequilibrium implies the
possibility of different, {\em mechanical} coupling of
reactions, upon which energy released in the center of one
reaction is directly transferred to the center of another
reaction. This fascinating possibility seems to be worth futher
studies.

\vspace{1pc}
\noindent
{\em Acknowledgements:} I wish to thank Manfred Eigen, Hans
Frauenfelder, No\-buhiro G\={o} and the late lamented Mikhail
Vladimirovich Volkenstein for, sometimes very heated,
discussions. A support of the Alexander von Humboldt Foundation
and the Polish State Committee for Scientific Research (project
2 P302 061 04) is acknowledged.

\setlength{\parindent}{0pc}
\vspace{2pc}
{\bf REFERENCES}

\vspace{1pc}
[1] Fersht, A. (1985) Enzyme Structure  and Mechanism, 2nd edn.,
Freeman, New York.

[2] Frauenfelder, H., Sligar, S. G. and Wolynes, P. G. (1991)
Science 254, 1598-1603.

[3] Kurzy\'{n}ski, M. (1994) Biophys. Chem., review article submitted.

[4] Kurzy\'{n}ski, M. (1993) FEBS Lett. 328, 221-224.

[5] Chernavsky, D. S., Khurgin, Yu. I. and Shnol, S. E. (1967)
Mol. Biol. (USSR) 1, 419-425; (1987) Biophysics (USSR) 32, 775-781.

[6] McClare, C. W. F. (1971) J. Theor. Biol. 30, 1-34.

[7] Blumenfeld, L. A. (1972) Biophysics (USSR) 17, 954-99;
(1981) Problems  of Biological Physics, Springer Series in
Synergetics vol. 7, Berlin.

[8] Williams, R. J. P. (1980) Chem. Soc. Revs. 9, 281-364;
(1993) Trends Biochem. Sci. 18, 115-117.

[9] Gavish, B. (1986) in: The Fluctuating Enzyme (Welch, G. R.,
ed.) pp. 263-339, Wiley, New York.

[10] Kov\'{a}\v{c}, L. C (1987) Curr. Topics Bioenerget. 15, 331-372.

[11] Shaitan, K. V. and Rubin, A. B. (1982) Mol. Biol. (USSR)
16, 1004-1009; (1982) Biophysics (USSR) 27, 386-390.

[12] Agmon, N. and Hopfield, J. J. (1983) J. Chem. Phys. 78,
6947-6959; (1983) 79, 2042-2053.

[13] Cartling, B. (1985) J. Chem. Phys. 83, 5231-5241.

[14] Kurzy\'{n}ski, M. (1994) J. Chem. Phys. 100, No. 13, in print.

[15] Clothia, C. and Lesk, A. M. (1985) Trends Biochem. Sci. 10,
116-120.

[16] Janin, J. and Wodak, S. J. (1983) Prog. Biophys. Molec.
Biol. 42, 21-78.

[17] Benett, W. S. and Huber, R. (1984) CRC Crit. Rev. Biochem.
15, 291-384.

[18] Haran, G., Haas, E., Szpikowska, B., K. and Mas, M., T.
(1992) Proc. Natl. Acad. Sci. USA 89, 11764-11768.

[19] Elber, R. and Karplus, M. (1987) Science 235, 318-321.

[20] G\={o}, N. and Noguti, T. (1989) Chemica Scripta 29A, 151-164.

[21] Rojewska, D. and Elber, R. (1990) Proteins 7, 265-279.

[22] Nishikawa, T. and G\={o}, N. (1987) Proteins 2, 308-329.

[23] Kidera, A. and G\={o}, N. (1990) Proc. Natl. Acad. Sci. USA
87, 3718-3722.

[24] Furois-Corbin, S., Smith, J. C. and Kneller, G. R. (1993)
Proteins, 16, 141-154.

[25] Kitao, A., Hirata, F. and G\={o}, N. (1991) Chem. Phys.
158, 447-472.

[26] McCammon, J. A., Gelin, B. R., Karplus, M. and Wolynes, P.
G. (1976) Nature 262, 325-326.

[27] Amadei, A., Linssen, A. B. M. and Berendsen, H. J. C.
(1993) Proteins 17, 412-425.

[28] Kurzy\'{n}ski, M. (1990) J. Chem. Phys. 93, 6793-6799.

\newpage
\setlength{\parindent}{0pc}
{\bf FIGURE CAPTIONS}

\vspace{1pc}
Figure 1. Schematic cross-section of the fundamental structural
unit of protein, a domain. Heavily shaded are solid-like
fragments of secondary structure ($\alpha$-helices or
$\beta$-pleated sheets) and weakly shaded are surrounding
liquid-like regions. Black is the catalytic centre localized at
two neighbouring solid-like elements. Models of Protein-Glass
type treat the dynamics of conformational transitions within the
native state of protein as a quasicontinuous diffusion of
structural defects through the liquid-like medium.
Alternatively, models of Protein-Machine type treat this
dynamics as a relative motion of solid-like elements, also of
the nature of a quasicontinuous diffusion, along a mechanical
coordinate, identified in this picture with the angle $x$. The
main assumption of the present paper is that the slow diffusion
dynamics controls the reaction. The picture can be reinterpreted
on a higher level of the hierachy: solid-like structural elements
represent then the whole domains moving in a multidomain
enzymatic macromolecule.

\vspace{1pc}
Figure 2. Single enzymatic reaction. (a) Conventional kinetics.
(b) Actual kinetics involving an astronomical number of
conformations of the enzyme or its complexes labelled with
indices {\em i} and {\em j}. (c) Protein-Machine model. Dynamics
of conformational transitions within each of the three chemical
species E, ER and EP is approximated by diffusion along a
mechanical coordinate. Perpendicular chemical transitions are in
general reversible; the arrows indicate only the assumed signs
of local reaction rates.

\vspace{1pc}
Figure 3. Protein-Machine model of a single enzymatic reaction.
Three different conformational potentials with minima at $x_0$,
$x_1$ and $x_2$ correspond to individual chemical states E, ER
and EP, respectively, of the enzyme. Three chemical transitions
are localized at the points 0, $x'$ and $x''$.

\vspace{1pc}
Figure 4. Diffusion time uphill, $\tau (0\!\to\! x)$, and
downhill, $\tau (x\!\!\to\! 0)$, the potential $x^2$.  Time is
counted in characteristic diffusion time units $\gamma^{-1}$.
Dimensionless coordinate $x$ is proportional to $\sqrt{kT}$.

\end{document}